\documentclass{eurovisDefinitions/egpubl}

\JournalSubmission    
\usepackage[T1]{fontenc}
\usepackage{eurovisDefinitions/dfadobe}  

\usepackage{cite}  
\BibtexOrBiblatex
\electronicVersion
\PrintedOrElectronic
\ifpdf \usepackage[pdftex]{graphicx} \pdfcompresslevel=9
\else \usepackage[dvips]{graphicx} \fi

\usepackage{eurovisDefinitions/egweblnk}

\graphicspath{{figures/}{pictures/}{images/}{./}} 

\usepackage{microtype}                 
\PassOptionsToPackage{warn}{textcomp}  
\usepackage{textcomp}                  
\usepackage{mathptmx}                  
\usepackage{times}                     
\usepackage{tabu}                      
\usepackage{booktabs}                  

\usepackage{url}
\usepackage{tabularx}
\usepackage{multirow}
\usepackage{amsmath}
\usepackage{amssymb}
\usepackage{nicematrix}
\usepackage{tikz}
\usepackage{array}
\usepackage{subfigure}

\newcommand{\squishlist}{
	\begin{list}{$\bullet$}
		{ \setlength{\itemsep}{0pt}      \setlength{\parsep}{3pt}
			\setlength{\topsep}{3pt}       \setlength{\partopsep}{0pt}
			\setlength{\leftmargin}{1.5em} \setlength{\labelwidth}{1em}
			\setlength{\labelsep}{0.5em} } }
	
	\newcommand{\squishlisttwo}{
		\begin{list}{$\bullet$}
			{ \setlength{\itemsep}{0pt}    \setlength{\parsep}{0pt}
				\setlength{\topsep}{0pt}     \setlength{\partopsep}{0pt}
				\setlength{\leftmargin}{2em} \setlength{\labelwidth}{1.5em}
				\setlength{\labelsep}{0.5em} } }
		
		\newcommand{\squishend}{
		\end{list}  }

\makeatletter
\renewcommand{\ps@plain}{%
\renewcommand{\@oddfoot}{\hfil\textrm{\thepage}\hfil}%
\renewcommand{\@evenfoot}{\@oddfoot}%
}
\renewcommand{\ps@empty}{%
\renewcommand{\@oddfoot}{\hfil\textrm{\thepage}\hfil}%
\renewcommand{\@evenfoot}{\@oddfoot}%
}
\makeatother

\title[VAKG]%
      {A Theoretical Approach for Structuring and Analysing Knowledge Provenance for Visual Analytics}

\author[L. Christino, S., Rezaeipour, E. Milios \& F. Paulovich]
{\parbox{\textwidth}{\centering L. Christino$^{1,2}$\orcid{0000-0002-8754-8460},
        S. Rezaeipour$^{1}$\orcid{0000-0002-2136-0713},
        E. Milios$^{1}$\orcid{0000-0001-5549-4675}
        and F. Paulovich$^{2}$\orcid{0000-0002-2316-760X} 
        }
        \\
{\parbox{\textwidth}{\centering $^1$Dalhousie University, Canada\\
         $^2$Eindhoven University of Technology (TU/e), Netherlands
       } 
}
}

\begin{document}

\teaser{
 \includegraphics[width=\linewidth]{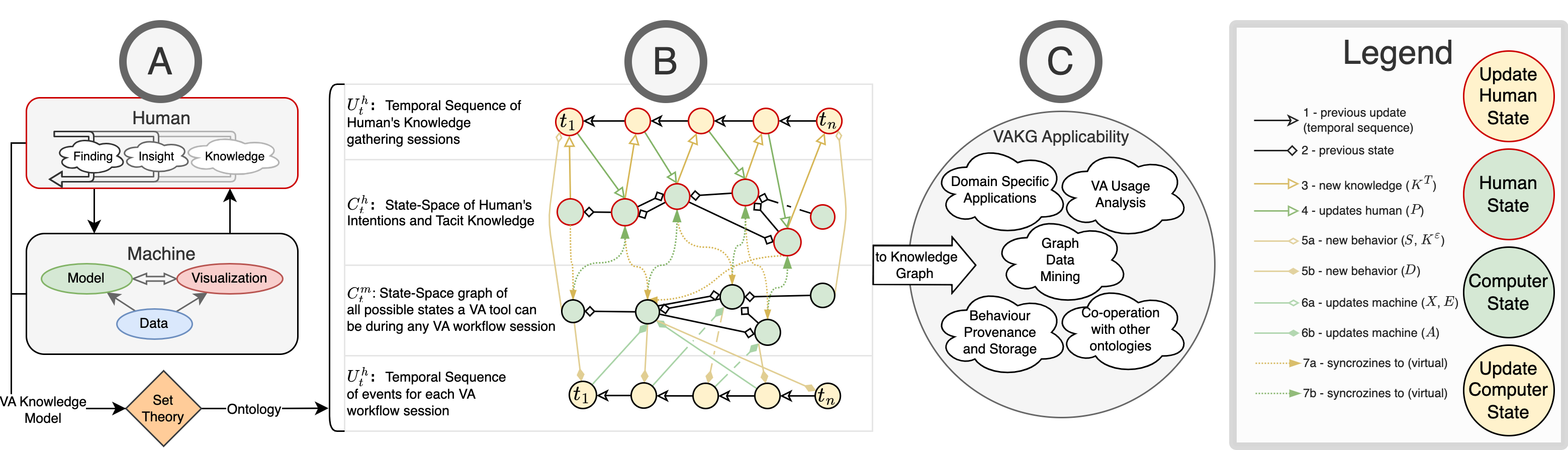}
 \centering
 \caption{VAKG unfolds the interactions within the current knowledge model (A) into a temporal knowledge graph (B), which is structured as a 4-way graph containing two temporal (green) and two static (yellow) knowledge graphs. By using VAKG, one can structure and store the user's knowledge-gathering process and all related interactions for eventual analysis (C).
 }
 \label{fig:teaser}
}

\maketitle
\begin{abstract}
  The primary goal of \textit{Visual Analytics (VA)} is to enable user-guided knowledge generation. Theoretical VA works to explain how the different aspects of a VA tool bring forth new insights through user interactivity, which itself can be captured through tracking methods for reproduction or evaluation. However, the process of automatically capturing the user's thought process, such as intent and insights, and associating it with user's interaction events are largely ignored. Also, two forms of interactivity capture are typically ambiguous and intermixed: the \textit{temporal} aspect, which indicates sequences of events, and the \textit{atemporal} aspect, which explains the workflow as sequences of states within a \textit{state-space}. In this work, we propose \textit{Visual Analytics Knowledge Graph (VAKG)}, a conceptual framework that brings VA modeling theory to practice through a novel \textit{Set-Theory} formalization of knowledge modeling. By extracting such a model from a VA tool, VAKG structures a 4-way \textit{temporal knowledge graph} that describes user behavior and its associated knowledge gain process. Such knowledge graphs can be populated manually or automatically during user analysis sessions, which can then be analyzed using graph analysis methods. VAKG is demonstrated by modeling and collecting Tableau and visual text-mining workflows, where comparative user satisfaction, tool efficacy, and overall workflow shortcomings can be extracted from the knowledge graph.


\end{abstract}

\section{Introduction}

Visual Analytics (VA) tools allow users to harness insights and knowledge from datasets~\cite{sacha2014knowledge}. By tracking this insight-generation process with provenance methods, like screen and mouse-click recording, researchers and industry alike can better understand the relationship between their tools and their users. The theoretical foundation of VA of Sacha et al.~\cite{sacha2014knowledge} proposes a ``knowledge generation model'' which not only discusses how the human and the machine interact during an insight-generation process but also discusses what elements within these interactions relate to the various data collected in provenance methods. Among existing works, Federico et al.~\cite{federico2017role} use the theory of Sacha et al.~\cite{sacha2014knowledge} to propose VA as a workflow of \textit{events}, which can potentially be associated with provenance concepts. Other works which propose novel VA systems also use the knowledge modeling theory to describe the process where their users gather knowledge, providing them with a basic best-practice design guideline of how to model~\cite{sacha2014knowledge, federico2017role}, structure~\cite{chen2017pathways, sacha2018vis4ml} and understand user behaviour~\cite{von2014interaction, mathisen2019insideinsights, heer2008graphical, xu2020survey}.

Structuring VA workflows has been a hot research topic~\cite{sacha2018vis4ml, chen2017pathways}, attracting enormous interest in tracking and analyzing user behavior to understand how knowledge is generated~\cite{xu2020survey}. In addition to investigating the role of the VA workflow during analysis, such works investigate how the users' pre-existing knowledge~\cite{heer2008graphical, battle2019characterizing} influence their experience during VA tasks. However, behavioral analytics do not yet use \textit{provenance} in light of the existing knowledge models or ontologies. Instead, each research endeavor develops its own method to acquire, structure, and analyze the users' knowledge-gathering process. Therefore, although Sacha et al.~\cite{sacha2014knowledge} describes the conceptual relationship between \textit{Machine} and \textit{Human}, some essential aspects are overlooked in behavior analysis. For instance, given that a VA tool follows the workflow of a specific VA model, how would one use this model as the means to acquire, structure, and store ongoing user interactions, or namely \textit{behavioral provenance}? Or how might the recorded data be analyzed to investigate and compare the knowledge gathered among several users, or namely \textit{knowledge provenance}? And how can a single dataset be defined which relates the user's behavior and the gathered knowledge where one can discover which sequences of actions lead to a new insight or which insights were attained due to using a specific, perhaps new, visualization? Or even how to use the answers to these questions for other downstream tasks, such as aiding the development of new tools or comparing different VA tools?

To address these open questions, we propose \textit{Visual Analytics Knowledge Graph (VAKG)}, a novel conceptual framework that proposes a formalized process to extract the underlying VA model of a VA tool, to design a knowledge graph ontology following the model, to define the data to be collected from the user behavior and knowledge gathering which fits said ontology, to populate a knowledge graph containing behavior and knowledge provenance data, and finally to use said knowledge graph for analysis of the relationship of behavior and knowledge. For this, we use existing VA knowledge models~\cite{sacha2014knowledge, federico2017role, van2005value} and reinterpret them as \textit{sets of information} and the process of how these sets interact. This way, VA is separated by its temporal aspect (e.g., temporal sequences of events versus atemporal \textit{state-spaces}) and ownership aspect (e.g., \textit{Human} versus \textit{Machine}). We then define a novel multi-layer knowledge graph structure that follows the \textit{sets of information} and their relationships. 

Our main contributions can be summarized as follows:

\squishlist
\item A reinterpretation of VA's knowledge model through \textit{Set-Theory} and the relationship between the modeled sets;
\item A domain-agnostic knowledge graph structure definition based on VA's knowledge model; and
\item A novel usage of a multi-layered Temporal Knowledge Graph architecture as a storage, analysis, and visualization mechanism of VA workflows for understanding the relationship between user behavior and knowledge acquisition.
\squishend

Consider this sample workflow: two data analysts~\cite{youtube1, youtube2} intend to investigate supermarket transactions dataset~\cite{globalsuperstore} using Tableau~\cite{murray2013tableau}. Their workflow can be summarized as downloading the dataset, verifying it is correct, and checking the store's profitability by creating and analyzing various visualizations. We propose that by mapping each step of the users' workflow to entities of a VA model~\cite{federico2017role}, VAKG provides a knowledge graph structure that relates user behavior and knowledge acquisition. Then, the knowledge graph can be built by recording each user's behavior and thought process. For instance, ``creating a profitability bar chart'' would be related to the next task of ``inspecting the tallest bar'' and the new knowledge of ``country X is the most profitable''. Finally, this knowledge graph can be used for downstream tasks using graph analytics. Questions like ``Which user had more insights during the process?'' or ``Which user took the least amount of time/steps to find the answer?'' can then be answered through the page-rank and shortest-paths algorithms, respectively. Therefore, VAKG not just models a workflow but also defines what data is relevant to be stored, such as user insights and interactions, in order to analyze the users' knowledge-gathering process, providing a unified and repeatable theoretical approach to bridge VA knowledge models, behavior provenance and knowledge provenance.

The remainder of the paper is structured as follows. In Sec.~\ref{sec:back} and Sec.~\ref{sec:related}, we introduce relevant concepts and discuss related work involving techniques that seek to formalize the VA knowledge flow, usages of knowledge graphs within VA, including how they differ from VAKG, and other concepts which tackle the ongoing knowledge evolution during data analysis. In Sec.~\ref{sec:methodology}, we extend the existing works of the theoretical knowledge model of VA to formalize VAKG. In Sec.~\ref{sec:applications}, we present possible applications of VAKG while comparing it with existing methods and justifications for further extending VAKG. We conducted a case study to demonstrate the practical application of the VAKG to a VA tool that analyses interactive clustering of textual documents in Sec.~\ref{sec:applications} called ModKT. The researchers developing ModKT are using the results to decide the next steps in their work. Finally, in Sec.~\ref{sec:future}, we discuss current limitations and the next steps within our research plan. In Sec.~\ref{sec:conclusion}, we draw our conclusions.


%
%

\section{Theoretical Background and Definitions}\label{sec:back}

Researchers typically prefer to define their workflow descriptively for particular use cases or follow certain well-tested processes. Theoretical research in the model design of VA workflows reflects this diversity very well. To properly position VAKG within the theoretical literature, we first define how the theoretical literature sets itself. Throughout this paper, we will follow the definitions of Chen et al.~\cite{chen2017pathways} where the contribution of theoretical VA works is categorized as one or more of the following:

\textbf{Principles and Guidelines}: Qualitative descriptions or rules which define a process that may lead to the desired outcome. Examples can be found in works that extract the qualitative elements of a VA workflow and define rules based on it~\cite {sacha2018vis4ml, brehmer2013multi}. 

\textbf{Taxonomy and Ontology}: A collection of concepts that defines a well-defined structure. Such research usually focuses on a novel theoretic ontology to structure the knowledge generation workflow~\cite{sacha2014knowledge, von2014interaction, sacha2018vis4ml, chen2020review, chen2019ontological, polowinski2013viso}.

\textbf{Conceptual models}: Abstract representation of a real-world process using a collection of theoretical taxonomies, typologies, and guidelines. For our purposes, a \textit{VA knowledge model} is a model of a user's knowledge generation throughout a VA process. Arguably the most prominent example of such a model is of \cite{sacha2014knowledge}. Generally speaking, knowledge modeling defines a workflow where insights lead to knowledge generation~\cite{andrienko2018viewing}.

\textbf{Theoretic frameworks}: Collection of operators which to measure a process (e.g., mathematical operators). For instance, the \textit{theoretic system} defined by Federico et al.~\cite{federico2017role} can describe and measure the process of many existing VA systems and tools.

\textbf{Quantitative laws}: Describes causal relationships between conceptual models by means of a theoretic framework. For example, Federico et al.~\cite{federico2017role} applies this concept when comparing multiple VA knowledge models.

\textbf{Theoretic systems}: An extension of a conceptual model which uses theoretic frameworks to define a real-world process formally. Federico et al.~\cite{federico2017role} extends several conceptual models in such a way as to formalize its methodology.

These concepts are not consistently used in the VA literature~\cite{federico2017role}. In order to better contextualize VAKG's goals, VAKG itself defines a theoretic system based on the set-theory theoretic framework, the conceptual model of Sacha et al.~\cite{sacha2014knowledge}, and the ontology of Federico et al.~\cite{federico2017role}. Beyond theory, we propose the practical use of VAKG by applying the proposed theoretic system in practice by performing behavior and knowledge provenance analytics. Because of this duality of VAKG, we classify it as a \textit{conceptual framework}. Nevertheless, the goal of VAKG was defined by investigating the connections between theoretical and practical related works. 

\section{Related Works}\label{sec:related}

This section presents an overview of how existing theoretical and non-theoretical works are related to VAKG while also considering the definitions of Sec.~\ref{sec:back}.

\subsection{Related Theoretical Works}

Knowledge modeling defines a workflow where user insights lead to knowledge generation~\cite{sacha2014knowledge, andrienko2018viewing}. For this, it defines the relationship between users' interactivity and all computer operations and data~\cite{sacha2014knowledge}. For instance, Fig.~\ref{fig:teaser}(A) summarizes this knowledge model, showing how knowledge generation and user interactivity are linked. Although such work is instrumental as a foundation throughout the VA literature, it cannot be directly applied in practice for provenance analysis.

On the other hand, ontology structures~\cite{sacha2014knowledge, von2014interaction, sacha2018vis4ml, chen2020review, chen2019ontological, polowinski2013viso} are being used as a means to link knowledge models to real-world workflows. Vis4ML~\cite{sacha2018vis4ml}, for instance, describes an ontology for machine learning in VA, and, with it, users can easily model and structure a machine learning workflow. Howsoever relevant these works may be for VAKG, their contribution is still only theoretical, not tackling how to store any data generated from executing a VA workflow nor discussing how or if such data can be collected and used for downstream tasks, such as data analysis. In other words, research on taxonomies and ontologies that structures knowledge gathering in VA does not, by design~\cite{chen2017pathways}, provide an overarching \textit{theoretic system} to link VA theory and the practice of provenance.

Since the origin of VA, significant work has been done to demonstrate the breadth and depth of \textit{knowledge} within VA~\cite{sacha2014knowledge}. The \textit{theoretic system} of Federico et al.~\cite{federico2017role} is versatile enough to describe many existing VA tools. More specifically, they show how the subsequent interactions and \textit{feedbacks} between the user and the computer are related. They also describe how automatic processes in data mining can generate new visualizations or how machine learning can help the user understand the data itself. Nevertheless, although the works listed and described by Federico et al.~\cite{federico2017role} may differ, VA's purpose of creating insight or knowledge through a given workflow is common to all of them and is generally done through interactivity between the user and computer~\cite{federico2017role, sacha2014knowledge, chen2019ontological}. Even though their \textit{theoretic system} can formalize the VA knowledge model and exemplify its application in practice, it by itself still lacks an \textit{ontology} to structure, store, and relate the provenance-related data, such as user behavior and the knowledge gathered.

Although the presented theoretical research, such as knowledge models, taxonomies, ontologies, and theoretic systems, are instrumental to understanding how current VA systems produce knowledge, we have also identified their insufficiency in providing insights into the ongoing knowledge generation process throughout a VA workflow. In other words, they cannot be used to simultaneously model, store, and create links between a VA tool's usage, the user's behavior during a VA workflow, and the user's knowledge-gathering process. VAKG attempts to bridge this gap. But our work does not try to redefine any of the taxonomies and principles described so far. Instead, VAKG uses the same taxonomies and principles as most~\cite{federico2017role, chen2019ontological, polowinski2013viso}. Also, although VAKG provides a more comprehensive structure to relate user behavior and the knowledge-gathering process, we recognize the existing works' advantage in other areas (e.g., data mining~\cite{sacha2018vis4ml} and machine learning~\cite{von2019informed, sacha2018vis4ml}). Therefore VAKG does not aim to supersede existing structures or ontologies with its own. Instead, VAKG requires that a given VA tool be modeled using existing VA models, then used to define its \textit{Knowledge Graph} structure. Thus, VAKG bridges the gap between VA theory and its applicability in practice to provide a cohesive structure to relate and analyze user behavior and knowledge gathering.

\subsection{Related Applications and Frameworks}

Theoretical research on VA's knowledge model has tackled the problem of knowledge gathering in many different ways. However, knowledge gathering within these works and systems is seen only as theoretical background. Federico et al.~\cite{federico2017role} lists many systems where a notable example is the work by Keim et al.~\cite{keim2010mastering}, which creates an application-specific knowledge-gathering process by utilizing automated analysis with human interaction; however, by verifying these related works, we note a lack of standardization of how to apply the theory in practice. Federico et al.~\cite{federico2017role} argues that since this knowledge-gathering loop is conceptual, it is ``often inconsistently used,'' which shows a missed opportunity to define how to apply such theory in practice in a consistent way. This inconsistency has another consequence: although their results relate to each other, these works do not seem to be able to communicate. In other words, we are unable to compare their results.

Furthermore, the two sides of knowledge gathering are often not well separated: the temporal sequence and the workflow's \textit{state space}, which denotes the set of all possible states independent of time. In other words, although a knowledge-gathering process can be defined as a linear sequence of new knowledge ``events'' over time, it can also be defined as a time-independent set of all gathered knowledge. With VAKG, we first explain the advantages of separating these concepts and the using each of the concepts in a unified framework. Different from other works~\cite{federico2017role}, VAKG uses this as one of its core design goals.

\textbf{User Behaviour Tracking and Behavior Provenance}: User-tracking and behavior analysis research has also been active~\cite{xu2020survey}. For instance, the user-tracking taxonomy of von Landesberger et al.~\cite{von2014interaction} models user behavior as a graph for analytical purposes. However, VA tools cannot integrate directly with theoretical works such as these. Instead, existing VA systems use these taxonomies as a theoretical or conceptual background while using the user-tracking data solely for specific domain use cases, as is extensively discussed by Xu et al.~\cite{xu2020survey}. For instance, the user's \textit{Tacit Knowledge}~\cite{federico2017role} is tracked in VA by many different feedback methods, such as manual feedback systems~\cite{bernard2017comparing, mathisen2019insideinsights}, manual annotations over visualizations~\cite{soares2019vista}, and inference methods that attempt to discover the user's insights by analyzing their interactivity patterns~\cite{nguyen2016sensemap, battle2019characterizing}. However, these works do not directly use any previously discussed theoretical results. Instead, they are only seen as a motivation for their domain-specific solutions. Among these VA systems, InsideInsights~\cite{mathisen2019insideinsights} and SenseMap~\cite{nguyen2016sensemap} are the only ones that get close to addressing this limitation. SenseMap first creates a graph network with behavior provenance, then allows users to analyze the recorded graph by manually constructing a so-called ``Knowledge Map''. InsideInsights, instead, records user behavior and user annotations simultaneously during the user's analytical process. Though InsideInsights and SenseMap provide a way to record and analyze user behavior, the proposed solutions are domain-specific and do not discuss the relationship between users' behavior and the knowledge gathered by the user. For instance, InsideInsights does not allow tracking auto-generated insights~\cite{spinner2019explainer} and does not account for automatic computer processes~\cite{federico2017role} or external agents~\cite{el2022biases, monadjemi2023human}. VAKG, however, also tackles these aspects.


\textbf{Knowledge Provenance}: Significant research has been done to better understand the concept and applicability of knowledge gathering in practice regarding \textit{knowledge provenance}. Knowledge provenance is a specialization of \textit{Data Provenance}~\cite{da2003knowledge, fujiwara2018concise} for collecting, storing, and tracking users' knowledge-related events. Knowledge provenance researchers argue that tracking user's knowledge gathering can be done by recording any change in the available datasets~\cite{da2009towards} (e.g., data pre-processing) or updates in visualizations~\cite{battle2019characterizing, xu2020survey, von2014interaction}. Among such works, Chang et al.~\cite{chang2009defining} attempt to use visual analysis within a Knowledge Base system, storing knowledge extracted from experts into a ``compressed'' format. Works such as these show examples of applying provenance to understand users' knowledge gathering.

Still, although these works describe ways to link knowledge gathering to user interactions, it is rare to see a differentiation between the temporal sequences of user-generated events and the atemporal \textit{state space} of the VA workflow. Therefore, the following two concepts are either merged or ambiguous in these works: the \textit{temporal} aspect, which indicates what and when users executed VA tasks, and the \textit{atemporal} aspect, which indicates what the possible VA workflow states and how they transition between each other are. Instead, when these works explicitly define a structure, they either store the temporal sequences of events without indicating whether they occurred previously or the state space without recording the temporal sequence of events. Similarly, these works assume that knowledge provenance is a subset of data provenance, or in other words, that all knowledge-related changes can be extracted from the user's behavior. This does not match with the knowledge definition of VA's knowledge models~\cite{sacha2014knowledge, federico2017role} where certain concepts, like behavior and knowledge, are separate. Likewise, most related works do not tackle how to interpret multi-user VA workflows~\cite{battle2019characterizing}, nor allow for comparisons between the user's exploratory space when compared to their motifs~\cite{xu2020survey}. VAKG bridges these gaps by modeling the difference between behavior and knowledge provenance and the difference between temporal events and temporal state space. VAKG encodes this model into a knowledge graph that relates users' behavior and knowledge-gathering sessions.

\textbf{Knowledge Graphs (KGs)}: While Knowledge Provenance focuses on tracking and storing knowledge, \textit{Knowledge Graphs} (KGs)~\cite{fensel2020introduction, chen2020review, li2023characterizing} have aimed to be a proper way to structure and analyze knowledge-related data. KG is a widely used technique to structure knowledge as a graph network, usually done by formalizing the structure as an ontology through the Web Ontology Language format~\cite{chen2019ontological, sacha2018vis4ml, von2019informed}. For instance, DBPedia~\cite{auer2007dbpedia} uses ontology design and KGs to transform unstructured knowledge into structured knowledge. In other words, KGs are a graph database of knowledge that employs knowledge model~\cite{sacha2014knowledge} ontologies. Compared to typical databases, the structure of KGs focuses less on the usual row-based structure~\cite{cashman2020cava} but uses the relationships between taxonomies as the foundation of knowledge. Although KG itself focuses on the structure of knowledge-related data, it is supported by various other graph-theory contributions, such as Graph Neural Networks (GNNs)~\cite{jin2020gnnvis}, graph visualizations~\cite{chang2016appgrouper, he2019aloha} and graph operations~\cite{ilievski2020kgtk}, like Page Rank and Traveling Salesman. KGs are, therefore, not limited to only providing a functional structure, but given a KG, users can employ graph analysis techniques to query and analyze the data. 

\textbf{Temporal Knowledge Graphs (TKGs)}: A notable sub-type of KGs is \textit{Temporal Knowledge Graphs (TKGs)}, where the graph edges encodes the temporal relationship of the data, such as ``order of events'' or ``time difference between events''~\cite{gottschalk2018eventkg}. That is, while a KG is a graph structure where knowledge reasoning is modeled as connections between classes or properties, such as ``George Washington \textit{is a} human'' and ``Canada \textit{is a} country'', a \textit{Temporal Knowledge Graph (TKG)} models these connections as the temporal relationship between the classes or properties. Many types of TKGs exist, and their temporal relationship varies among them. For instance, TKGs can relate two nodes by temporal co-occurrence. An example of such a KG would be all purchases done between different businesses within a supply chain, where the product ``Mayonese'' may have been bought by ``Walmart'' from the seller ``Hellmann's'' on ``25/06''. In this TKG, the connection between the three nodes: Walmart, Hellmann's, and Mayonnaise, would be ``25/06''. Though some existing works which define knowledge graphs~\cite{cashman2020cava, xu2020survey, nguyen2020knowledge, li2023characterizing} or ontologies~\cite{sacha2018vis4ml, chen2019ontological} are already used for structuring knowledge and behavior provenance, no current work, as far as the authors know, uses TKGs to structure knowledge provenance.

\textbf{Process Mining}: The act of structuring and analyzing a process in a graph format has been extensively researched by Process Mining~\cite{van2004process}. Indeed, the relationship between Process Mining and VA has grown tremendously in recent years. Process Mining proposes a way to define any given process by a workflow consisting of nodes and their relationships. The concepts of \textit{events} and \textit{knowledge graphs}, which are very relevant for VAKG, have appeared in many recent works~\cite{fahland2022process}, showing how Process Mining is a proven form of modeling processes for provenance purposes~\cite{zeng2011method, van2015extracting}. Yet, Process Mining is centered on behavior and events, that is, behavior provenance. VAKG aims to relate behavior to \textit{knowledge generation}, which differs from existing works' goals.

\subsection{Survey Compilation and Goals}

We compiled a survey analyzing the most relevant literature cited so far to verify how the theoretical concepts are applied in practical related works. We found that the main differentiation of VAKG is its theoretically grounded pipeline, which, in a simplified manner, one must: model a given VA tool using a VA model and ontology~\cite{federico2017role}, declare a knowledge graph structure that matches said ontology, perform data collection through behavior and knowledge provenance to populate the knowledge graph, and finally analyze said knowledge graph (see Sec.~\ref{sec:methodology}). Thus, VAKG's major goals are:

\begin{itemize}
    \item \textbf{G1. Analysis-centric VA Model:} Temporal and atemporal interpretations of \textit{Human} and \textit{Machine} components of the VA workflow are used, but inconsistently, so VAKG proposes a consistent one which partitions the VA workflow as:
    \begin{itemize}
	    \item \textbf{G1.1:} Temporal-sequences of user's knowledge gathering (Knowledge Provenance or \textit{Human Updates})~\cite{sacha2014knowledge, sacha2018vis4ml, polowinski2013viso, von2014interaction, federico2017role, brehmer2013multi, von2019informed, chen2020review, jin2019recurrent, da2009towards, battle2019characterizing, clifton2012advanced, mathisen2019insideinsights, bernard2017comparing};
        \item \textbf{G1.2:} User intentions and insights which occur within a VA workflow (\textit{Human State-Space} or just Human State)~\cite{sacha2014knowledge, sacha2018vis4ml, polowinski2013viso, federico2017role, brehmer2013multi, von2019informed, chen2019ontological, chen2020review, auer2007dbpedia, chang2016appgrouper, he2019aloha, jin2019recurrent, battle2019characterizing, heer2008graphical, clifton2012advanced, mathisen2019insideinsights};
        \item \textbf{G1.3:} The VA tool's states during all VA workflows are modified due to user behavior (\textit{Machine State-Space} or just Machine State)~\cite{sacha2014knowledge, sacha2018vis4ml, polowinski2013viso, von2014interaction, brehmer2013multi, chen2019ontological, chen2020review, auer2007dbpedia, chang2016appgrouper, he2019aloha, jin2019recurrent, callahan2006vistrails, da2009towards, battle2019characterizing, heer2008graphical, clifton2012advanced, spinner2019explainer};
        \item \textbf{G1.4:} Temporal-sequences of the VA tool events/tasks which are executed during VA workflow sessions (Behaviour Provenance or \textit{Machine Updates})~\cite{sacha2014knowledge, sacha2018vis4ml, polowinski2013viso, federico2017role, brehmer2013multi, auer2007dbpedia, jin2019recurrent, callahan2006vistrails, da2009towards, battle2019characterizing, spinner2019explainer, mathisen2019insideinsights, bernard2017comparing};
    \end{itemize}
    \item \textbf{G2. Ontology of the VA Workflow:} Formalization of a structure that, while being rooted in an existing VA knowledge model~\cite{sacha2014knowledge, federico2017role}, describes the VA workflow following \textbf{G1}~\cite{sacha2018vis4ml, polowinski2013viso, brehmer2013multi, von2019informed, chen2019ontological, chen2020review, auer2007dbpedia, chang2016appgrouper, he2019aloha, jin2019recurrent, da2009towards};
    \item \textbf{G3. Data Retention:} The structure is used as a schema of a data retention solution where to collect and store user behaviors and interactions during a VA workflow~\cite{polowinski2013viso, von2014interaction, federico2017role, chen2019ontological, chen2020review, auer2007dbpedia, chang2016appgrouper, he2019aloha, jin2019recurrent, callahan2006vistrails, battle2019characterizing, heer2008graphical, spinner2019explainer, mathisen2019insideinsights, bernard2017comparing};
    \item \textbf{G4. Data Analysis Capabilities:} Use the data and/or structure to perform analysis, such as per-user analysis, user comparison, usage comparison, and so on~\cite{von2019informed, chen2019ontological, chen2020review, auer2007dbpedia, chang2016appgrouper, he2019aloha, jin2019recurrent, callahan2006vistrails, da2009towards, heer2008graphical, clifton2012advanced, spinner2019explainer, bernard2017comparing};
\end{itemize}

The next section describes how VAKG reaches these goals.

\section{The VAKG Conceptual Framework}\label{sec:methodology}

Let's assume a group of researchers created a VA tool for the analysis of temporal series and now wants to understand if, how, and what users learn while using their tool. The \textit{Visual Analytics Knowledge Graph (VAKG)} method gives this group a formalized process to extract the underlying VA model of a VA tool, design a knowledge graph that follows the model, and define which data from the user needs to be collected for a thorough provenance of the user's behavior when using the tool and their newly acquired knowledge from the tool.

First, VAKG requires that a VA knowledge model is matched to the tool~\cite{sacha2014knowledge, federico2017role} (see Fig.~\ref{fig:teaser}[A]). By VAKG reinterpretation of the model in the lens of \textit{Set Theory} (\textbf{G1}), VAKG identifies what are the unique elements that constitute a VA workflow of that specific VA tool and what are their relationships to each other. This VA workflow is then structured following VAKG's \textit{ontology} that relates the users' interaction events and knowledge generation (see Fig.~\ref{fig:teaser}(B) and \textbf{G2}). The result is a knowledge graph structure that separates the workflow's temporal aspect, which is defined as behavior sequences of events (\textbf{G1.4}) and knowledge-gathering sequences of events (\textbf{G1.1}), the workflow's atemporal aspect, which is structured as the VA tool's \textit{state-space}(\textbf{G1.3}) and the users' knowledge \textit{state-space} (\textbf{G1.2}). VAKG then uses the knowledge graph structure as the design pattern for a multi-layer \textit{Temporal Knowledge Graph (TKG)} where the VA tool can record user sessions (\textbf{G3}). Finally, this populated knowledge graph is available to users, such as the research group of the example above, to apply graph-network techniques to analyze, predict, and recommend user behavior and knowledge-gathering effectiveness when using the tool (\textbf{G4}).

\subsection{Foundation: VA Knowledge Model and Set Theory Reinterpretation}\label{sec:formal}

The theoretical background of VA's knowledge model is a foundation work for research within VA (see Sec.~\ref{sec:related}). Unlike such works, we use the knowledge model of Sacha et al.~\cite{sacha2014knowledge} as a foundation to formalize and derive VAKG (\textbf{G1}). This section reinterprets the VA knowledge model to define the four aspects of our Analysis-centric VA Model: human update, human state, machine state, and machine update.

\begin{figure}[tb]
    \centering
    \includegraphics[width=1\columnwidth]{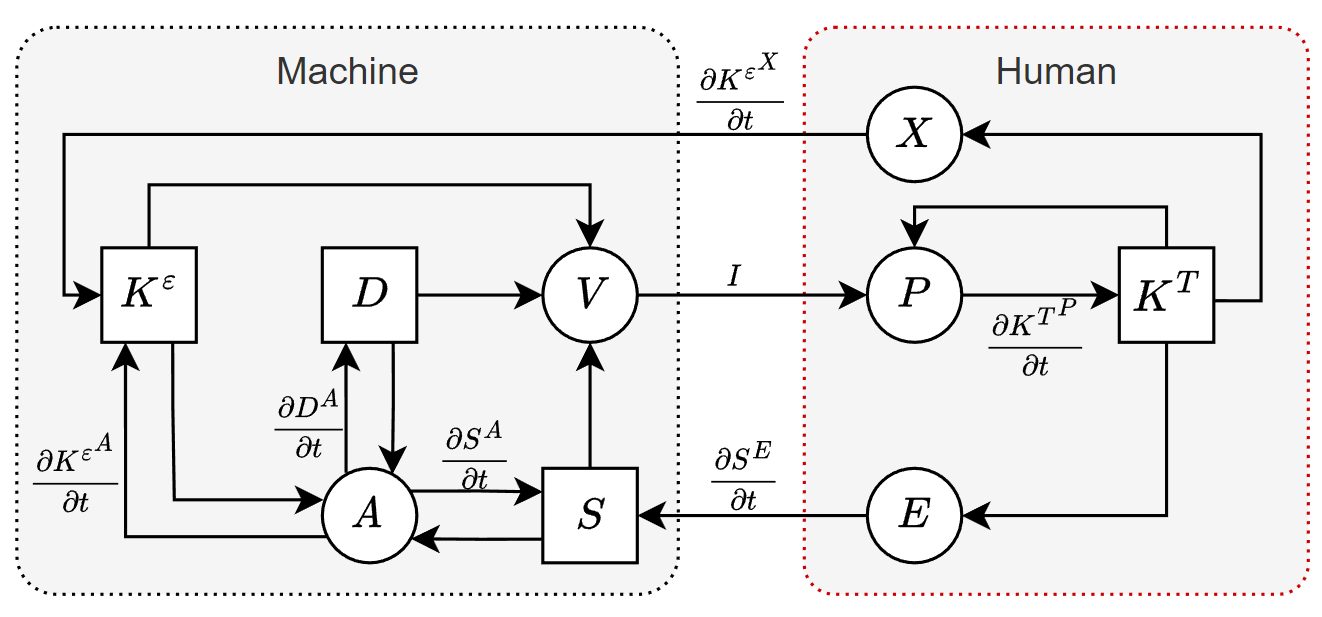}
    \caption{Conceptual Model of Knowledge-Assisted VA~\cite{federico2017role}. VAKG structures its mathematical framework by deriving the equations from this model.}
    \label{fig:knowledge1}
\end{figure}

The simplistic representation of VA's knowledge model shown in Fig.~\ref{fig:teaser}(A) characterizes its two main actors: \textit{Humans} and \textit{Machines}. This concept originates from the literature where knowledge is generated over time~\cite{sacha2014knowledge} though the interaction between Human and Machine~\cite{federico2017role}. The literature also proposes a mathematical interpretation of the VA model called the ``Conceptual Model of Knowledge-Assisted VA'' as the foundation of the knowledge model, which is expressed visually in Fig.~\ref{fig:knowledge1}. 

All in all, the VA interactivity model is divided between two separate actors (machine and human) and describes how knowledge is generated, converted, and used within the VA discourse. Each actor is then associated with a taxonomy of available actions (1): analysis $A$, visualization $V$, externalization $X$, perception/cognition $P$, and exploration $E$. These actions are connected by intermediate stateful taxonomies (2): explicit knowledge $K^\epsilon$, data $D$, specification $S$, and tacit knowledge $K^T$; and (3) a non-persistent artifact: image $I$'' (see Fig.~\ref{fig:knowledge1}).

From Fig.~\ref{fig:knowledge1}, we find that the circle nodes $\{V, P, E, X, A\}$ represent elements that cause changes within a VA tool. For instance, the visualization \textbf{V} resides in the machine space, and it causes changes in the perception/cognition of the user \textbf{P}, providing new insights. Similarly, the exploration task \textbf{E}, which is in the human space, executed by the user can update the VA tool's specification \textbf{S}, which may update the data or visualization being shown within the VA tool. From Fig.~\ref{fig:knowledge1}, we also find that the set of rectangle nodes $\{K^T, D, S, K^\epsilon\}$ represents static information. For example, elements such as data \textbf{D}, specification \textbf{S}, and tacit knowledge \textbf{Kt} represent the fact that the VA workflow has static information about a dataset, the state of the VA tool, and the user's tacit knowledge, respectively. 

From Federico et al.~\cite{federico2017role}, we can identify all the moving parts within the knowledge model iterative loop of a given VA tool. The first contribution of VAKG is to reinterpret the iterative loop of Fig.~\ref{fig:knowledge1} through the lens of \textit{Set Theory}, allowing this process to be applied to other VA models and tools. Therefore, first, we define four sets of information: the \textit{Machine Update} $U^m_t = \{V_t, A_t\}$, the \textit{Machine State} $S^m_{t} = \{D_{t}, S_{t}, K^{\epsilon}_{t+1}\}$, the \textit{Human Update} $U^h_t = \{X_t, P_t, E_t\}$, and the \textit{Human State} $S^h_{t} = \{K^T_{t}\}$. Next, from Fig.~\ref{fig:knowledge1}, we extract how each of these elements relates to each other. Each equation below represents which information (rectangle node) directly depends on a process (circle nodes) and which processes directly depend on information: 

\begin{align}
\label{eq:firstknowledge1}
K^T_{t+1} &\Leftarrow P_{t+1} \Leftarrow K^T_{t} + V_{t}(I)\\
\label{eq:firstknowledge2}
K^\epsilon_{t+1} &\Leftarrow X_{t+1} + A_{t+1} \Leftarrow K^T_{t} + (K^\epsilon_{t} + S_{t} + D_{t})\\
\label{eq:firstknowledge3}
S_{t+1} &\Leftarrow E_{t+1} + A_{t+1} \Leftarrow K^T_{t} + (K^\epsilon_{t} + S_{t} + D_{t})\\
\label{eq:firstknowledge4}
D_{t+1} &\Leftarrow A_{t+1} \Leftarrow K^\epsilon_{t} + S_{t} + D_{t}
\end{align}

By using these equations, we reach that the human state and machine state are updated as follows:
\begin{align}
\label{eq:CH}
S^h_{t+1} =& \{K^T_{t+1}\} \Leftarrow U^h_{t+1}(S^m_t) \\
\label{eq:CM}
S^m_{t+1} =& \{D_{t+1}, S_{t+1}, K^{\epsilon}_{t+1}\} \Leftarrow U^m_{t+1}(S^m_t) + U^h_{t+1}(S^h_t)
\end{align}

That is, the \textit{human state} is updated due to a \textit{human update} caused by some change within the \textit{machine state} (Eq.~\ref{eq:CH}). Similarly, the machine state is updated due to a human or machine state change (Eq.~\ref{eq:CM}).

Following this process, any VA tool can be decomposed into the four sets of state and process entities, and the equation list with the relationships between the entities within the sets. Back to the example scenario discussed in the introduction: a data analyst wishes to investigate the supermarket dataset~\cite{globalsuperstore} using Tableau~\cite{murray2013tableau}. In this simplified scenario, the ``VA tool'' is tableau. The available usages of Tableau can be mapped to the nodes of Fig.~\ref{fig:knowledge1}. For example, let's assume a user wants to create a visualization in Tableau. The data $D$ is the supermarket dataset, the state of tableau $S$ represents what visualization, if any, is currently being shown, and the creation of a new visualization $E$ would update the state of tableau $S$, generating the new visualization $V$ with which the user can investigate $P$. In other words, the node $E$, part of the Human Update set, leads to a new visualization. In mathematical terms: $S_{t+1} \Leftarrow E_{t+1} \equiv S^m_{t+1} \Leftarrow U^h_{t+1}$. That is, in this example a human update $U^h_{t+1}$ led to a new machine state $S^m_{t+1}$. Still, no new data $D$ has been generated yet, so it was removed from the equation. 

Now, if the user discovers a new insight $K^T$ from the visualization and adds it as a custom text or annotation $X$ to the visualization, new explicit knowledge $K^\epsilon$ would be saved into the tool, causing subsequent updates following the equations above. We see, therefore, that the equations above are helpful not just to define how each of the processes $\{V, P, E, X, A\}$ updates the static information $\{K^T, D, S, K^\epsilon\}$, but to define how these updates can simultaneously be understood by its ownership (machine or human) and by its timing.

\subsection{VAKG Ontology and Knowledge Graph Definition}\label{sec:ontology}

So far, we have described VAKG's foundation through its four aspects: human update, human state, machine state, and machine update. We also described how each aspect interacts with the others through set equations. Yet, to store data of the users' knowledge generation process, VAKG defines a \textit{Knowledge Graph} (KG) structure where its nodes and relationships correspond to the four aspects of VAKG and their update relationship according to the set equations. This structure allows VAKG to use existing graph databases directly, unlike the domain-specific VA ontologies designed recently~\cite{sacha2018vis4ml, von2019informed, chen2019ontological}. The final structure is exemplified in Fig.~\ref{fig:teaser}(C), where the four color-coded horizontal lanes display each of the four aspects. 

\begin{figure}
    \centering
    \includegraphics[width=.75\columnwidth]{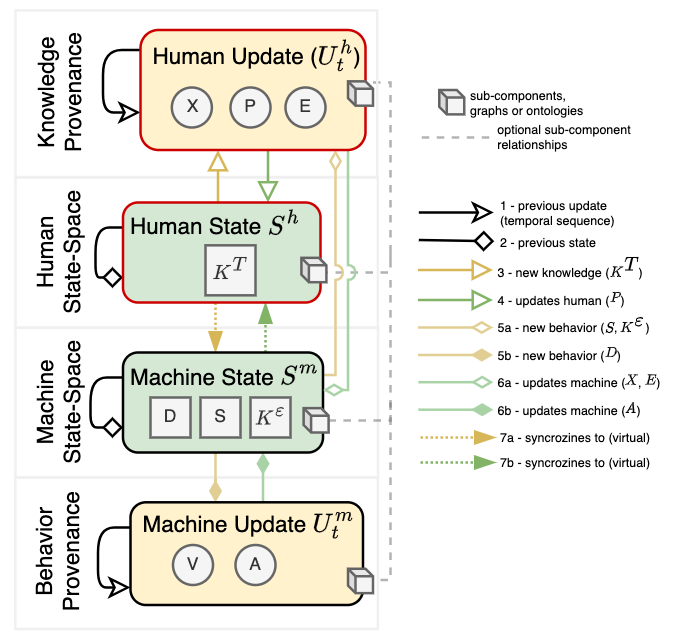}
    \caption{VAKG ontological design. The four different lanes of VAKG are represented. Two are KGs that describe the possible states of the inner property maps or sub-graphs. Two are TKGs describing the sequences of updates, such as the sequence of insights and knowledge gathering by users in a VA workflow or the sequence of computer events. }
    \label{fig:entiredesign}
\end{figure}

By following the Web Ontology Language (OWL)~\cite{chen2017pathways}, VAKG divides the space in two ways: by its ownership (human or machine) and by its timing (state or update), which defines the four ontology classes: Human-Update, Human-State, Machine-State, and Machine-Update. From Eqs.~\ref{eq:CH} and~\ref{eq:CM}, VAKG defines the relationships between the four classes, which are represented in Fig.~\ref{fig:entiredesign}. Namely, the relationship links [1] and [2] found in Fig.~\ref{fig:entiredesign} relate the previous human/machine state/update to the current one, [3] and [4] represent Eq.~\ref{eq:CH} where a change in $K^T$ leads to an update in $P$, and [5] and [6] similarly represent Eq.~\ref{eq:CM}. Finally, VAKG defines two extra relationships [7], synchronizing the two state spaces. This way, if a change in specification (e.g., new visualization) causes the user to perceive something (through [5a]), leading to new knowledge (through [3]), VAKG relates the starting \textit{machine state} and ending \textit{human state} through [7b]. Similarly, if this new knowledge leads the user to externalize [6a] (e.g., add text to the visualization) ending in a new explicit knowledge $K^\epsilon$, VAKG relates the starting \textit{human state} and ending \textit{machine state} though [7a]. Fig.~\ref{fig:teaser}(B) exemplifies a simple knowledge graph following VAKG's ontology.

\subsubsection{VAKG Property Map and Data Collection Guideline}\label{sec:propertymap}

An integral part of our proposal is to record users executing a VA workflow and to enable its usage for analysis. This process is called \textit{provenance} (see Sec.~\ref{sec:related}). While the usual way of thinking of \textit{Knowledge Graph (KG)} is to focus on \textit{classes} and their \textit{relationships}, VAKG instead gives significant importance to \textit{property-maps} (also called \textit{class properties} or \textit{data properties}). Property-maps employ the idea that every \textit{class} can contain attached data. In VAKG, the property-maps of the four classes of nodes are expected to contain the relevant information of that specific class. For instance, in Fig.~\ref{fig:entiredesign}, we see that the class human state should contain the information related to the user's tacit knowledge $K^T$, and the machine state information related to the dataset $D$, specification $S$, and explicit knowledge $K^\epsilon$. However, the \textit{property-map} design pattern is interchangeable with the other common design patterns~\cite{myroshnichenko2009mapping}, which removes any perceived limitation of our approach.


VAKG, therefore, records the information of a node as a property-map, but how should it be recorded? And what information \textit{exactly} should be included? This question is the underlying reason for our descriptive formalization (see Sec~\ref{sec:formal}) because, without it, we would not know precisely what information should be stored in each of the node's property-maps. For instance, we have previously described how a machine state would store information related to the dataset $D$, specification $S$, and explicit knowledge $K^\epsilon$, but how much of such information should be stored? Although theoretically, one could argue that storing all information related to a given state is the solution.

It is not reasonable to expect that the usage of VAKG would necessarily require such an amount of information. Therefore, we propose that the property-map of any \textit{State} should, at the very least, uniquely identify that specific \textit{State} within the entire state-space of VAKG. Similarly, the property-map of any \textit{Update} should uniquely identify the changes between the two Machine or two Human \textit{States}, including the timestamp of when the change occurred. This definition establishes that a given Machine or Human \textit{State} can repeat if the same condition occurs multiple times. 
It is important to note that since each specific use case of VAKG may vary, this part of VAKG is treated as a \textit{design guideline}.

Therefore, it is essential to note that the center two lanes of Fig.~\ref{fig:teaser}(B) and Fig.~\ref{fig:entiredesign} are \textit{atemporal} because their connection is not temporally dependent. In other words, \textit{Machine} and \textit{Human} states are related not through temporal dependency but through their transition relationship. Structures like finite-state machines and discrete-time Markov chains also use atemporal transition relationships similar to VAKG. For example, a machine state is related to a human state through Fig.~\ref{fig:entiredesign}[7b] if that machine state $S^m$ caused the human state $S^h$ to leave a prior state $S^h_a$ and reaches another $S^h_b$. This may also be read as ``$S^h_a$ lead to $S^h_b$ when $S^m$ happened'' where the word ``when'' does not refer to ``exact time'' but to the idea of ``consequence'' instead.

This way, by repeating an earlier example, if a change in specification (e.g., new visualization) within a machine state $S^m$ causes the user to perceive something (through [5a]), leading to new knowledge (through [3]) and consequently a new human state $S^h_b$, VAKG relates the starting \textit{machine state} $S^m$ and ending \textit{human state} $S^h_b$ through [7b]. VAKG also links the two human states by relationship [2], as shown in Fig.~\ref{fig:entiredesign}. Note that the same process happens when a new human state leads the machine state $S^m_a$ to change to a new state $S^m_b$.

A consequence of this structure is that nodes in the machine and human space-states which are close (e.g., low number of relationships between the nodes) indicate that these nodes are similar since one state can quickly be reached from another through a low number of ``updates''. Also, if two machine states or two human states are directly connected, only a single update is responsible. 
\subsection{VAKG in Practice}
\label{sec:applications}

The running example used until now involves two Tableau users verifying and analyzing a global supermarket store. In this section, we expand on this example as a use case of VAKG. We also discuss another use case with a VA tool called ModKT~\cite{rezaeipourfarsangi2022interactive}.


\subsubsection{Tableau Use-Case}

\begin{figure}[ht]
    \centering
    \subfigure[User~\cite{youtube1} downloads and analyses the data by validating if the data is correct, then finds an insight with a choropleth map, and finally validates it through a bar chart.]{\includegraphics[width=0.45\textwidth]{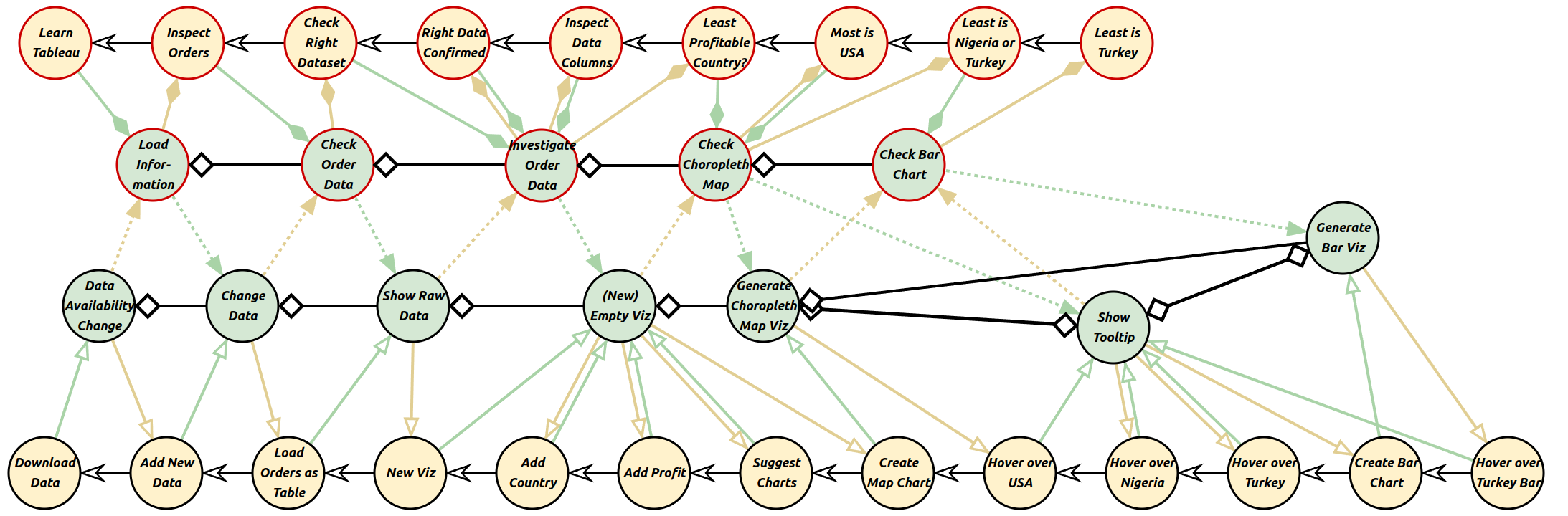}\label{fig:example1}} 
    \subfigure[User~\cite{youtube2} downloads but does not validate the data. He then builds step-by-step a specific choropleth map design to forward to management without discussing any insight.]{\includegraphics[width=0.45\textwidth]{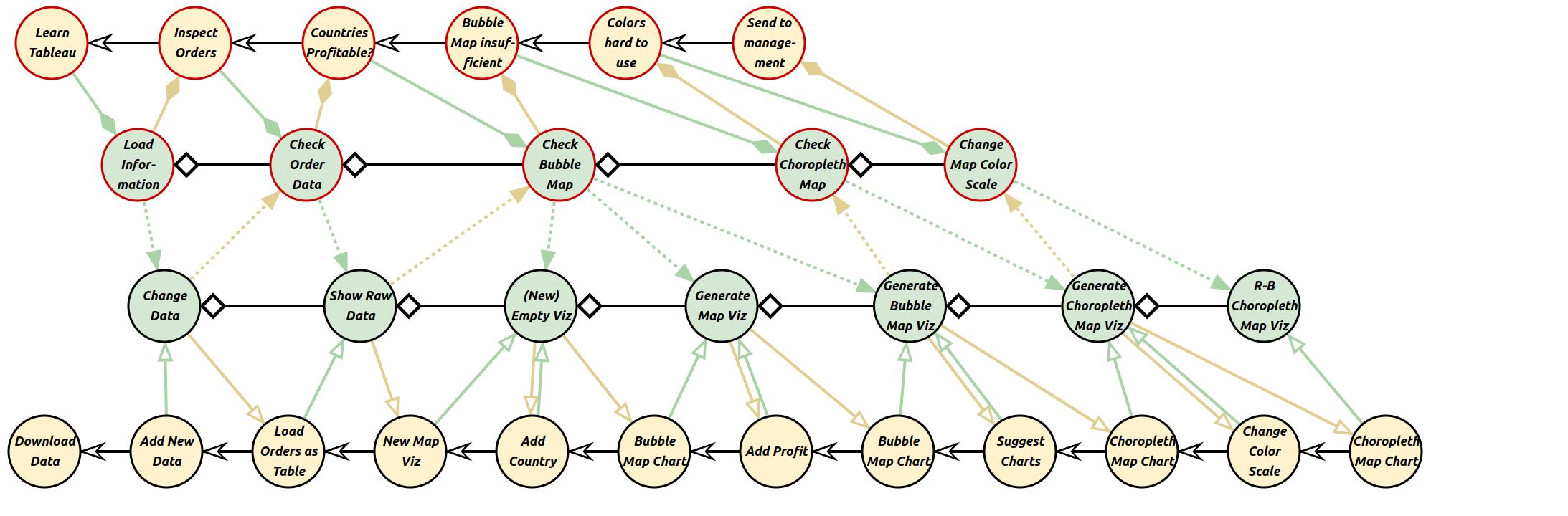}\label{fig:example2}} 
    \caption{VAKG of users performing visual analysis of a global superstore's profitability. The two graph networks are shown separately for better readability. Still, all \textit{green} nodes (\textit{state-space} nodes) with the same name are a single node in the VAKG graph, which is composed of both graphs simultaneously where the \textit{state-space} (green nodes) connects to the individual user's sequence of events (yellow nodes). }
    \label{fig:example}
\end{figure}

The first use case to discuss is the running tableau example where two~\cite{youtube1, youtube2} data analysts investigate the supermarket dataset~\cite{globalsuperstore} using Tableau~\cite{murray2013tableau}. By watching the two videos, we can extract a list of tasks, interactions, questions, and insights that each user did. For brevity, here is a small sample of these insights: ``task: download data'', ``task: find least profitable country'', ``interaction: create new visualization'', ``interaction: hover over the visualization'', and ``insight: the least profitable country is $C$''. Each process step can be mapped to one of $\{V, P, E, X, A, K^T, D, S, K^\epsilon\}$ from Fig.~\ref{fig:knowledge1}. For instance, ``download data'' is a change to the data $D$, ``create a new visualization'' changes the specification $S$, and ``found least profitable country'' is a perception process $P$ resulting in new knowledge $K^T$. By mapping all users' steps to the proper taxonomy, VAKG defines what data of each step one may need, such as the modified data in $D$, the new visualization type in $S$, and the new insight in $K^T$. VAKG also classifies each workflow step as machine update, machine state, human state, and human update (see Fig.~\ref{fig:entiredesign}). For instance, a data change is a new machine state, and a new insight is a new human state. Similarly, VAKG associates the sequence of actions, such as the act of looking at the visualization $P$ is the human update that led to the new insight $K^T$ (see Fig.~\ref{fig:knowledge1}). After applying VAKG to all steps, the result is the knowledge graph seen in Fig.~\ref{fig:example} of the videos' content~\cite{youtube1, youtube2}.


\subsubsection{ModKT Use-Case}\label{sec:case_study2}

We also apply VAKG to ModKT~\cite{cabral2020visual, rezaeipourfarsangi2022interactive}, an interactive clustering VA framework, to investigate which features of the tool are being used, how effective the features appear to be given insights gained while using ModKT, and to surface relevant next steps of its authors' research. In this section, we describe the tool, how VAKG was applied to it, and some preliminary results extracted from informal usage of the tool. This example demonstrates how VAKG can be applied to more complex VA tools and workflows. 

\begin{figure}[tb]
    \centering
    \includegraphics[width=.99\columnwidth]{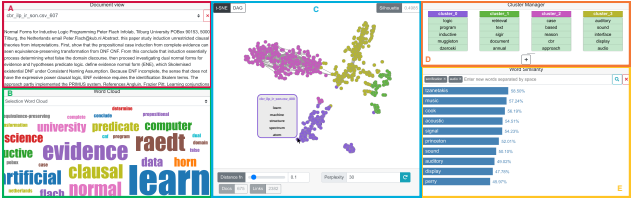}
    \caption{The modular architecture interface provides a glimpse into its functionality, operating on a collection of 660 documents related to computer science subjects. The interface includes several components: A Document view (A) that shows the content of a selected document; a Word cloud view (B) that presents either the focus document or a cluster in a visual form; a Graph view (C) that illustrates the similarity relationships between the documents in the corpus; a Cluster Manager (D) allowing users to examine clusters and provide feedback to the clustering algorithm; and a Word similarity view (E) which presents a bar chart indicating the similarity between user-provided query words and the most similar identified words.}
    \label{fig:Mod_KT}
\end{figure}

ModKT is a tool that ingests a set of documents, such as research articles, extracts key terms of each document, and applies key-term-based clustering~\cite{sherkat2018interactive} to the corpus. ModKT uses the articles' metadata, such as abstract, authors, title, journal, bibliography type, publication year and month, and URL, for clustering. Users can visualize each document through Word Clouds, the corpus of documents through dimensionality reduction, and the comparison of the extracted key terms to custom user-defined words. Users can customize the parameters for clustering and dimensionality reduction to discover sets of (dis)similar documents and visually analyze their (dis)similarities. An overview of the system is presented in Fig.~\ref{fig:Mod_KT}.

For this user study, we have set up ModKT with a list of 660 scientific articles in the computer science field covering various text-mining visualization subjects. In order to apply VAKG to it, we follow the methodology process of Sec.~\ref{sec:methodology}: model the VA tool, structure the knowledge graph, perform provenance to store user sessions with the tool, and analyze the resulting knowledge graph.

Due to the data and interactions used and expected by ModKT, we notice that even though the VA knowledge model of Federico et al.~\cite{federico2017role} (see Fig.\ref{fig:knowledge1}) could be used, it has elements that are not used by the tool, differing from examples given so far. For instance, ModKT does not allow externalizing knowledge $X$ into new explicit knowledge $K^\epsilon$. 

\begin{figure}[tb]
    \centering
    \includegraphics[width=1\columnwidth]{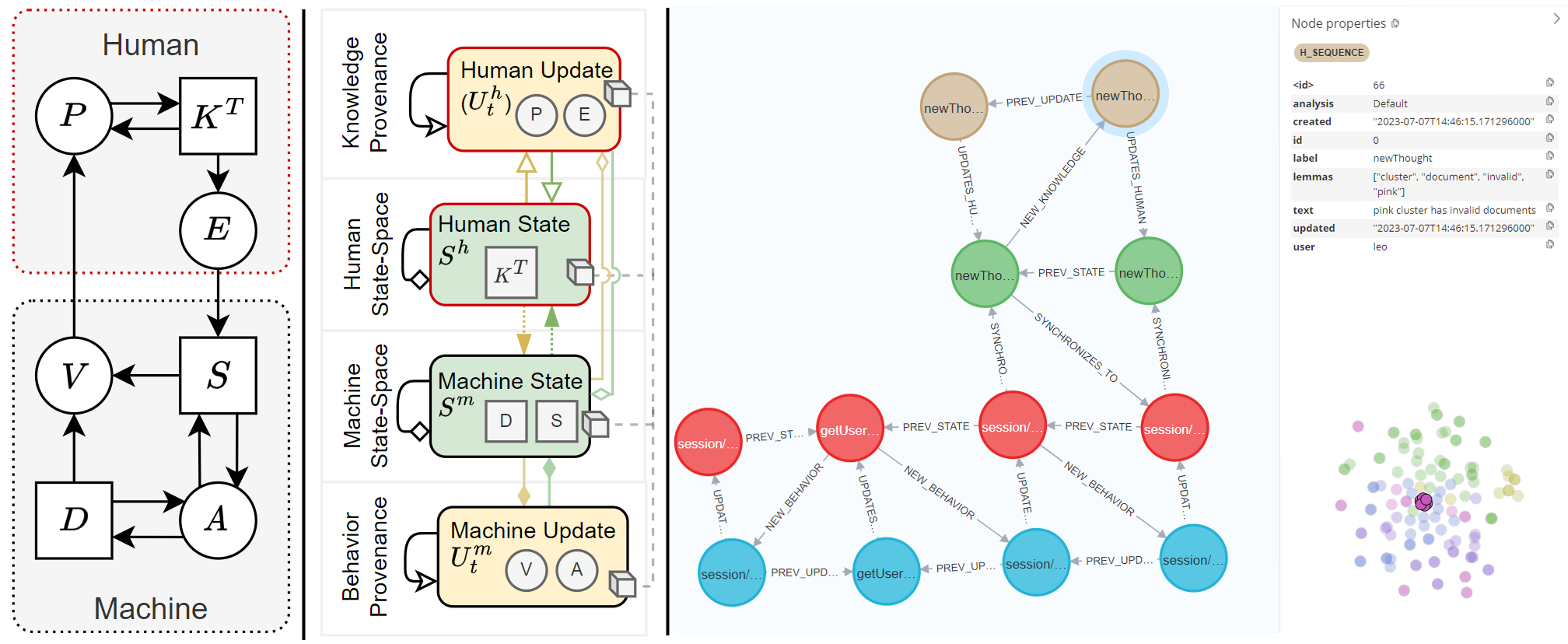}
    \caption{VA knowledge model of ModKT [Left], its corresponding VAKG structure [Center], and a knowledge graph generated using such structure [Right]. The model and structure only differ from Fig~\ref{fig:knowledge1} due to the removal of elements $X$ and $K^\epsilon$. The knowledge graph shows four interactions and two user thoughts, the last of which is highlighted and displayed on the right panel as ``Pink cluster has invalid documents''. This thought was provided by the user when the pink-colored documents were investigated, as shown by the ModKT dimensionality reduction visualization on the bottom right.}
    \label{fig:modktmodel}
\end{figure}

So far, VAKG has been exemplified only on one of the available VA models~\cite{federico2017role}. However, we can apply the same VAKG \textit{theoretic framework} to other VA models by following the same procedure of applying a set Theory reinterpretation to the VA model and extracting the VAKG ontology out of the equations. In the case of ModKT, let's consider that its VA model follows Fig.~\ref{fig:modktmodel}[Left]. The difference between this new model and the one used in Sec.~\ref{sec:methodology} is the absence of $X$, $K^\epsilon$ and any relationships that either $X$, $K^\epsilon$ had with any other entity of the model. Following our framework, the VAKG ontology is shown in Fig.~\ref{fig:modktmodel}[Center].

Next, we apply provenance to the ModKT tool. For this, we developed a sample implementation of this VAKG structure~\cite{leonardo_christino_2023_8124221} that receives an API call from ModKT at every user interaction. This sample implementation collects and populates a knowledge graph with user behavior data $\{V, S, D, A\}$, which is collected from mouse interactivity, and user thoughts $\{P, E, K^T\}$, which are collected by asking the user to type or speak into the microphone. Using speech-to-text and natural language processing, we extract keywords from the user's text and associate text and keywords with the user's behavior at the time. One of the collected user behavior and thought processes is shown in Fig.~\ref{fig:modktmodel}[Right].



\begin{figure}[tb]
    \centering
    \includegraphics[width=1\columnwidth]{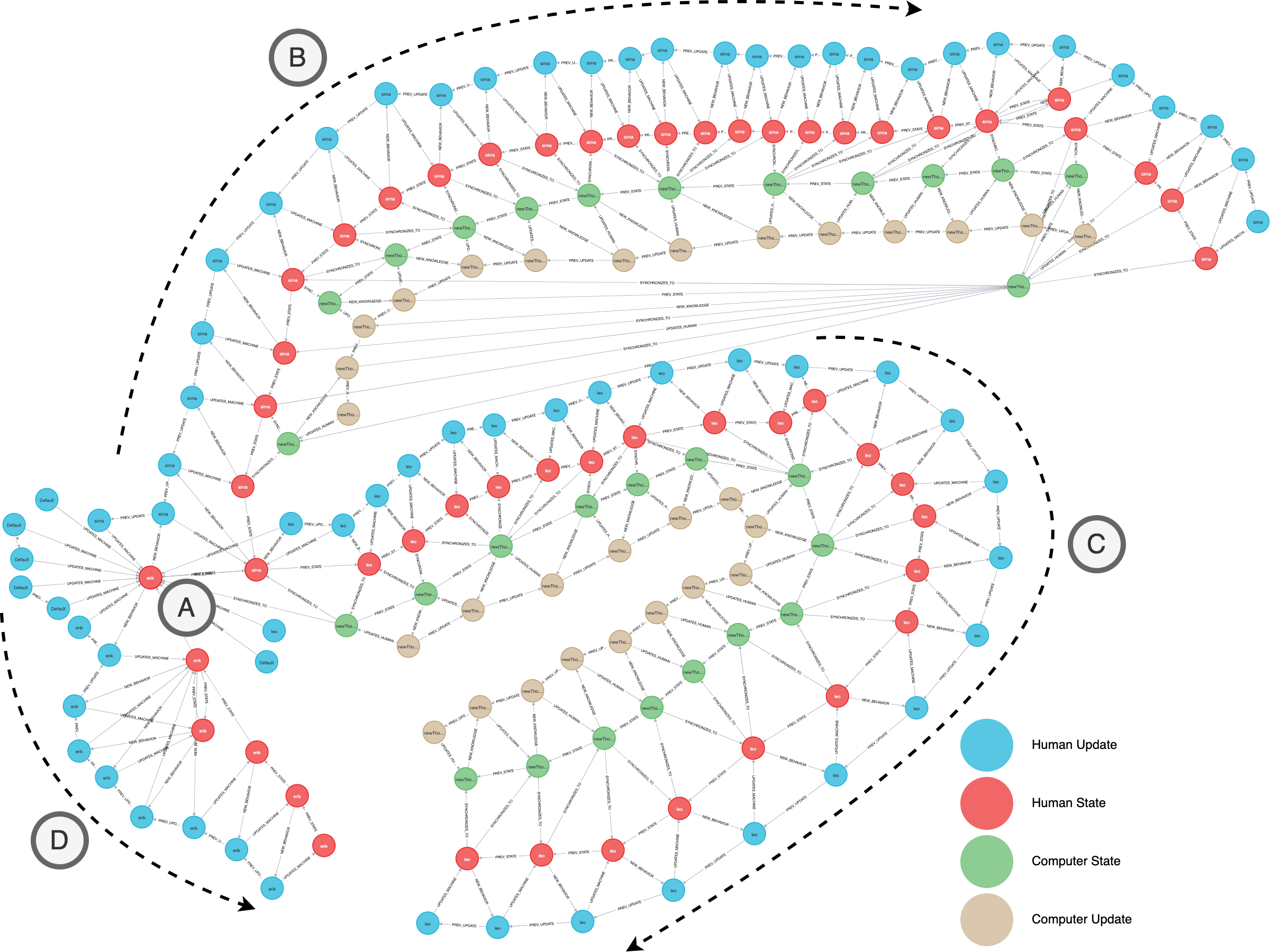}
    \caption{VAKG generated from three users interacting with ModKT. All users start at [A]. Users B and C had the same first step before diverging. User B performed two investigations, one with many steps [top] and a short one [bottom], ending in the same machine state. User D provided no thoughts, but the collected interactivity showed a back-and-forth interactivity pattern before concluding.}
    \label{fig:modktvakg}
\end{figure}

Next, we requested three researchers from our lab to use ModKT with VAKG. They were given 400 documents with abstracts and titles and were tasked with finding visualization-related articles that could be included in their next research article. With the resulting knowledge graph, we could understand how the tool is generally used, list several shortcomings of the tool, and compare how the users differ in their process. The full resulting VAKG knowledge graph is shown in Fig.~\ref{fig:modktvakg}. 

With a knowledge graph generated, we can now explore the graph through the node-link diagram of Fig.~\ref{fig:modktvakg}. Although all three users (B, C, and D) started out with a T-SNE projected visualization, user [C] immediately changed to a force-based layout because of the large amount of overlap, which was included in his Human Update ``T-SNE not useful, switching to DAG''. Though user [C] started by interacting with the visualization, user [B], instead, started by changing the clustering parameters, aiming to create a cluster to show documents related to visualization. Although user [B] could create such a cluster, their process was thwarted because the vast majority of the abstracts found were focused on NLP research and not visualization. 

Interactivity-wise, the graph shows that although users took different approaches. To analyze common patterns among users, we could investigate the graph through the visualization, but for better scalability, we opted to run graph queries~\cite{noauthororeditorneo4j} to fetch certain information. For instance, by querying for the nodes where the users used the forced-based layout (DAG), we can see that all three users used the forced-based layout (DAG) and changed its parameters at some point. Also, by fetching which documents were clicked by each user, all users were shown to have clicked on some of the documents to read more through the abstract and word cloud view. That said, all users were also shown not to have been very successful in finding visualization-related abstracts, which indicates that the issue was not the users nor the tool but the insufficient number of documents loaded into the tool.

Feedback related to the tool functionality was also collected from the users. They discussed topics specific to ModKT, such as layout problems, the aforementioned T-SNE overlap problem, a less-than-ideal experience when reading the abstracts, and little usefulness of the word cloud. VAKG also collected indirect feedback on the tool's functionality. For instance, using simple counting and the aforementioned Page-Rank algorithms, we queried the number of state nodes visited by multiple users, but it was very small. That is, the three users had nearly no overlap in their interactivity, showing that the search space of the tool is vast, likely too vast. The tool's researchers concluded that reducing the possible interactivity and replacing text-only panels with static visualizations is a potentially good next step for the tool. This was corroborated by counting the number of interactions the users had until they reached certain conclusions.

Though VAKG, ModKT researchers could analyze user exploration paths, check which features of ModKT were most and least used, check which clustering parameters were used, and collect much feedback for future steps. ModKT researchers claim to have gained insights into the tool's capabilities and limitations by visualizing and analyzing the workflow of the individual users, giving them valuable insights into the next step of their research.

While this process could have been done through surveys, thinking-aloud sessions, screen recording, and other manual techniques, the entire process of collection and structuring was done automatically by VAKG. Indeed, ModKT researchers praised VAKG, indicating that future user studies of their tool would be able to be done in a much more automatic and scalable manner. What previously would involve planning and manual labor, now users of ModKT just had to do interactions in an unsupervised environment and write or speak their thoughts into the built-in text widget added to ModKT. The sample implementation provided~\cite{leonardo_christino_2023_8124221} collected and populated the knowledge graph shown in Fig.~\ref{fig:modktvakg}, which was then analyzed to reach the conclusions above. 

\subsubsection{Using VAKG for Analysis}

\begin{figure}[tb]
    \centering
    \includegraphics[width=.8\columnwidth]{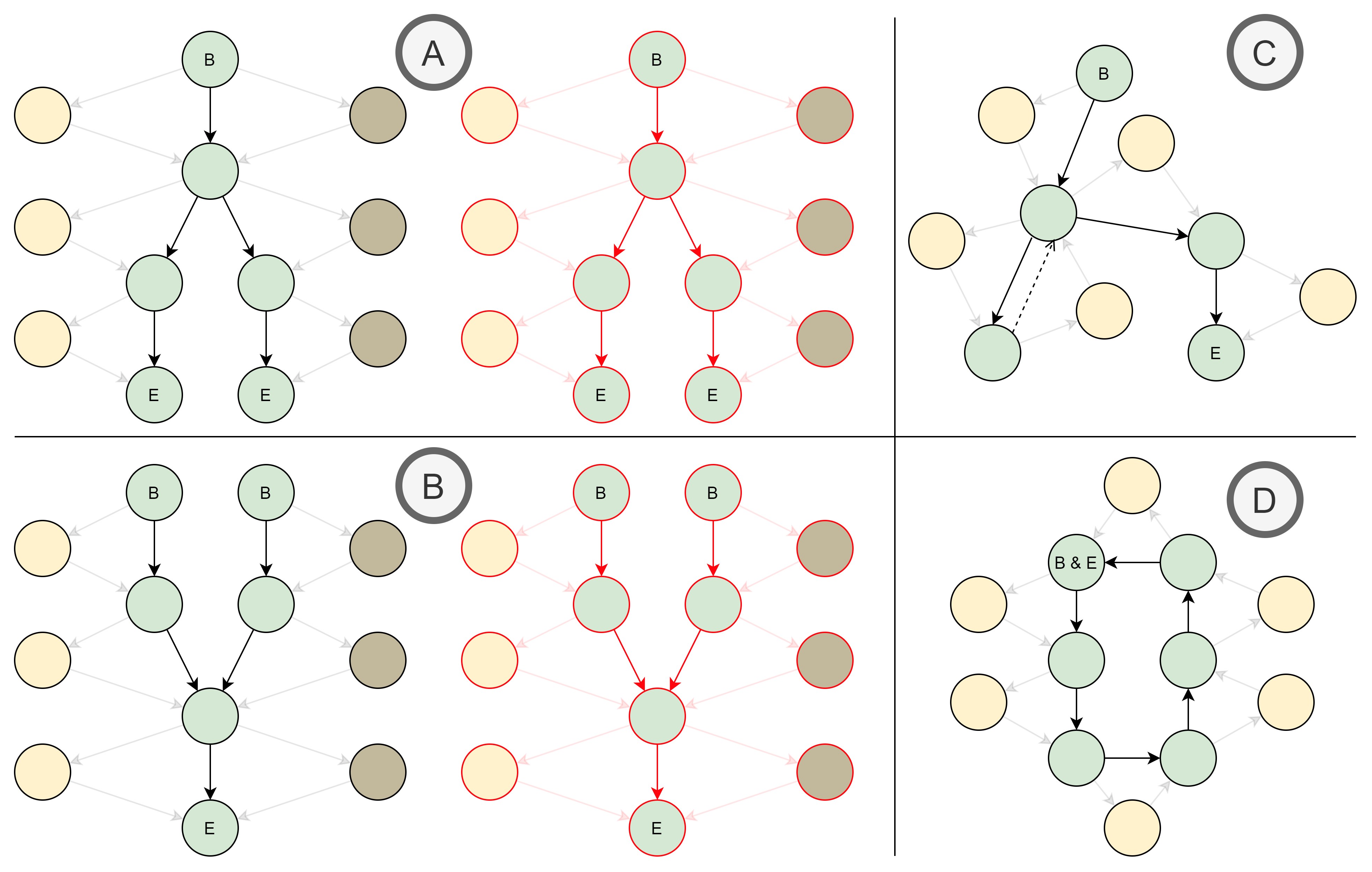}
    \caption{VAKG examples of graph patterns where B is ``beginning'' and E is ``End''. In [A], the two users began together but diverged at a certain point, and in [B], they started with different tasks but eventually converged. In [C], users backtrack using an ``undo'' operation, and in [D], users loop to a prior state.
    }
    \label{fig:vakgexamples}
\end{figure}

VAKG's structure allows users to leverage their existing techniques to perform analysis. The VAKG of Fig.~\ref{fig:vakgexamples} displays features that can occur in a VAKG graph and their respective meaning. Fig.~\ref{fig:vakgexamples} shows two users performing a workflow with diverging paths [A], converging paths [B], backtrack [C], and loop [D]. The example of [D] can also be seen in the Tableau example of Fig.~\ref{fig:example1} when the user interacts with a map visualization through tooltips. Knowing this, we can apply graph analysis techniques to VAKG to answer questions. 

One of the most ubiquitous techniques for graph network analysis is PageRank~\cite{gleich2015pagerank}, where it is possible to extract and rank graph nodes based on the number of their connections to other nodes. We can, for instance, extract the users' most ``important'' state by applying PageRank over VAKG's \textit{Machine-States}. By applying PageRank of the \textit{Machine-States} of the Tableau example (see Fig.~\ref{fig:example}), we find that user 1 interacted with tooltips more than any other interaction. PageRank also reveals that this specific machine state has the highest amount of \textit{update} relationships among all \textit{Machine-States}. Now, if we consider the full VAKG where both graphs of Fig.~\ref{fig:example} are merged, then PageRank of all relationships indicates that the node ``New Empty Viz'' is the most visited node with 19 connections. By filtering the connections by user, we see that this result is mainly due to the first user, who has $8$ connections to his \textit{Machine-Update} timeline versus $4$ of the second user. The same analysis can be done from the perspective of the \textit{Human-States} to discover that the ``Check Choropleth Map'' node is the most connected, which leads us to conclude that the users gathered more insights from the choropleth maps than any other visualizations. It is important to note that although these results can be checked visually in Fig.~\ref{fig:example}, in examples with hundreds of users where each performs hundreds of interactions, the use of such a ranking algorithm becomes significantly more important. 

Other graph network and knowledge graph techniques and tools~\cite{han2014chronos, ilievski2020kgtk, cashman2020cava, noauthororeditorneo4j, wang2018ripplenet} can also be applied. A cycle detection algorithm~\cite{qiu2018real} can find any closed cycles within VAKG. By applying it to the same Tableau example, we find that user 2 has no loops within his \textit{Machine-States}, which means that he never retraced his steps (see Fig.~\ref{fig:vakgexamples}[D]). By applying a shortest-path algorithm~\cite{madkour2017survey} over the \textit{Human-Update} nodes, we also find that the second user had fewer knowledge-related events, such as insights or questions, than the first, information which can aid in investigating if the tool is properly achieving its goals. We can also apply graph summarization~\cite{fujiwara2018concise} to simplify large graphs, apply KG completion~\cite{li2018embeddingvis} to analyze whether its users explored all the features of a VA tool, explore KGs through other tools~\cite{cashman2020cava}, and analyze the KG by its embedding~\cite{wang2017knowledge}.

We also extend the same examples to analyze users' workflows while performing tasks with different tools. For instance, assuming a third user performs a similar workflow to users 1 and 2~\cite{youtube1, youtube2} but with a different tool, such as PowerBI~\cite{ferrari2016introducing}. Graph analysis through PageRank, shortest-path, and other previously discussed techniques can again be used to compare how well the two tools performed. Indeed, if VA tools shared a VAKG of their user evaluation, other researchers would be able to download them and add new data from their own users and/or tools, allowing such researchers to compare their users or tools to existing state-of-the-art tools and past users using techniques like PageRank and shortest path to demonstrate the effectiveness of their new tool in terms of knowledge gathering effectiveness, which can potentially be used to transform the way VA research discusses and discloses user evaluation.

\subsection{VAKG Evaluation Discussion}

We aimed to follow existing theoretical work's example-based evaluation. However, due to the novelty of VAKG as a conceptual framework with modeling, ontology, structuring, and analysis components, no other work, as far as the authors know, can directly compare. That said, VAKG does not aim to supersede any specialized work in their areas. Instead, VAKG uses related work as its foundation. VAKG can also be easily extended by other works, potentially adding more data to VAKG's property-maps as sub-components.

For instance, here we compare VAKG to Vis4ML~\cite{sacha2018vis4ml}, which focuses on proposing an ontology for ML-related tasks. Compared to Vis4ML, VAKG focuses on a different aspect of the VA workflow: the classification of VA taxonomy based on ownership (Human and Machine) and temporality (state-space or process) and the relationship between them. VAKG proposes a knowledge graph structure and a methodology for populating the knowledge graph. Vis4ML only proposes a structure with no direct application to define or populate a knowledge graph or to define how to analyze the resulting data. Indeed, by comparing VAKG to all other related work, VAKG stands out as the only one that proposes a methodology to structure a given VA tool's model as a knowledge graph that can perform knowledge and behavior provenance. 
Similar results are found when comparing VAKG to other theoretical-focused works~\cite{chen2019ontological, sacha2014knowledge, sacha2016analytic, federico2017role}, which are here omitted due to space constraints.

When comparing VAKG to the results of practical related works, we find that VAKG is uniquely positioned to provide a comprehensive knowledge graph for their required behavior and user-knowledge analysis requirements. However, it is important to note that this comparison is limited because VAKG is a conceptual framework. For instance, InsideInsights' results~\cite{mathisen2019insideinsights} show that allowing users to visualize the VA workflows of certain analysis processes is highly beneficial through interviews and usage scenarios. In our MobKT user case (see Sec~\ref{sec:case_study2}), we confirm that visualizing the resulting knowledge graph is useful. Yet, the results of our work show that VAKG provides a structure for analyzing behavior and knowledge provenance, as opposed to InsideInsights' report of user behavior. Indeed, in practice, most works focus on behavior analysis~\cite{von2019informed, chen2019ontological, chen2020review, auer2007dbpedia, chang2016appgrouper, he2019aloha, jin2019recurrent, callahan2006vistrails, da2009towards, heer2008graphical, clifton2012advanced, spinner2019explainer, bernard2017comparing}. VAKG is novel in its inclusion of knowledge provenance as part of the resulting knowledge graph. 

\section{Limitations and Future Work}
\label{sec:future}

When comparing to other ontologies, it is essential to note that VAKG's focus is not on its descriptive power~\cite{sacha2018vis4ml}, but on its ability to model and structure the user's knowledge gain process. Therefore, VAKG does not solve the issue of how to perform user-tracking~\cite{mathisen2019insideinsights}. VAKG also does not expand the analytical arsenal of user behavior or provenance techniques~\cite{battle2019characterizing, da2009towards}, but provides a novel structure that is optimized for the use of said techniques for various analytical use cases. In future work, we aim to investigate the best approaches for user-tracking, behavior/knowledge provenience, and knowledge graph analysis when applying VAKG in domain-specific use cases. If required, we might contribute novel approaches. We also plan to investigate semi-automatic or automatic provenance techniques to assist the applicability of VAKG.

Although VAKG has focused on defining a property-map way to store the knowledge-gathering process, other works have proposed other methods as well~\cite{sacha2018vis4ml}. That said, knowledge graphs are not limited to a single structure at a time, as is the nature of graph data, so it is easy to imagine that two different knowledge graphs could co-exist. Therefore, although we argued that VAKG's structure is more capable than other existing ontologies, we recognize that this is mainly because the resulting knowledge graph can be extended, allowing others to use different ontologies or models as part of VAKG through custom property-maps or by linking VAKG nodes to a totally separate custom knowledge graphs. However, we believe that this integration needs to be addressed separately in domain-specific frameworks or application use cases. Results and evaluation of these future works will also be driven by their use cases, which do not fit within the contribution presented in this paper. Since existing ontologies~\cite{federico2017role, sacha2018vis4ml, curry2020real, xu2020survey, von2014interaction} can then co-exist with VAKG, we plan for future work to explore possible combinations of related work's ontologies as future domain-specific contributions. 

We have experienced that VAKG can quickly result in large and complex KGs, which are hard to visualize and may cause issues related to storage space if used indiscriminately. So far, we have provided examples that were simple enough to be explained and visualized. Still, we attempted to store dozens of user workflows as a VAKG, and the result was too complex to visualize. Indeed, we recognize that the complexity depends on the modeled and recorded workflow, though graph network analysis is always possible. We plan on investigating better ways to visualize both simple and complex VAKGs, especially when considering what analysis is being done as future work. 

The most critical limitation of VAKG, perhaps, is that user-tracking has been broadly seen negatively. User protection laws and initiatives, like Europe's General Data Protection Regulation (GDPR)~\cite{EUdataregulations2018} and Apple's ``Ask not to track'' features, are just a few examples. Although VAKG is not a novel way to perform user-tracking, user consent for tracked and behavior analysis is undoubtedly a relevant concern. However, this concern is not new and is shared by all related works which tackle user-tracking or behavior analysis. We also argue that in many cases, the users of VAKG are the same whose behavior is being tracked, which means that they probably would accept and welcome the necessary tracking since they would do the analysis. Further study is needed to analyze how impactful this would be.


%
%

%
%


\section{Conclusion}\label{sec:conclusion}

We have presented VAKG, a conceptual framework to structure a given VA tool as a 4-way temporal knowledge graph that describes user behavior and knowledge gathering during the execution of a VA workflow. We propose that by modeling a VA tool with VAKG, we obtain a knowledge graph structure that captures the required substances from user knowledge-gathering sessions. Users then populate the knowledge graph with behavior events, such as interactions, and knowledge events, such as intents and insights. Then, the knowledge graph can be used to analyze user behavior, the knowledge-gathering process, and the interactive relationship between the two. The resulting knowledge graph is by design standardized across users and tools, allowing for graph-based analytics of domain-specific processes (e.g., EDA), usage patterns, and user knowledge gain performance among multi-user and multi-tool scenarios.

In practice, VAKG's resulting graph represents an overview of the VA workflows' usage and the collective experiences and knowledge generated by their users. VAKG is extensible and adaptable to various situations and domains, including its extension to incorporate other models or ontologies. Using VAKG as a provenance architecture, the generated knowledge graph can also be analyzed through existing graph-analytics techniques, such as visualizations, shortest path analysis, and page ranking. We applied VAKG to two examples: data analysis with Tableau and ModKT~\cite{rezaeipourfarsangi2022interactive}, and discussed how the resulting knowledge graph allows us to better understand the path taken by the user to reach new knowledge, how users differ in their experience of seeking knowledge, and which parts of the tool were most and least used, among other results. When compared to existing works, VAKG was shown to be unique in its approach in bringing VA model theory into practice for behavior and knowledge provenance tasks.



\section*{Acknowledgment}
The authors acknowledge the support of the Natural Sciences and Engineering Research Council of Canada (NSERC).


\bibliographystyle{eurovisDefinitions/eg-alpha-doi}
\bibliography{main}

\end{document}